\documentclass[aps,pra,showpacs,twocolumn]{revtex4}
\usepackage{amsmath}
\usepackage{graphicx}
\usepackage{float}
\usepackage[ansinew]{inputenc}
\usepackage{array}
\usepackage{color}
\usepackage{amsxtra}
\usepackage{amstext}
\usepackage{amssymb}
\usepackage{latexsym}
\usepackage{gensymb}
\usepackage{dsfont}
\usepackage{braket}
\begin{document}

\title{Spin squeezing an ultracold molecule}

\author{M. Bhattacharya}
\affiliation{School of Physics and Astronomy, Rochester Institute of Technology, 84 Lomb Memorial Drive,
Rochester, NY 14623, USA}

\date{\today}
\begin{abstract}
In this article we present a concrete proposal for spin squeezing the ultracold ground state polar
paramagnetic molecule OH, a system currently under fine control in the laboratory. In contrast to
existing work, we consider a single, non-interacting molecule with angular momentum greater
than $1/2$. Starting from an experimentally relevant effective Hamiltonian, we identify a parameter regime
where different combinations of static electric and magnetic fields can be used to realize the single-axis
twisting Hamiltonian of Kitagawa and Ueda [M. Kitagawa and M. Ueda, Phys. Rev. A {\bf 47}, 5138 (1993)],
the uniform field Hamiltonian proposed by Law et al. [C. K. Law, H. T Ng and P. T. Leung, Phys. Rev. A
{\bf 63}, 055601 (2001)], and a model of field propagation in a Kerr medium considered by Agarwal and
Puri [G. S. Agarwal and R. R. Puri, Phys. Rev. A {\bf 39}, 2969 (1989)]. To support our conclusions,
we provide analytical expressions as well as numerical calculations, including optimization of field
strengths and accounting for the effects of field misalignment. Our results have consequences for
applications such as precision spectroscopy, techniques such as magnetometry, and stereochemical
effects such as the orientation-to-alignment transition.
\end{abstract}

\maketitle
\section{Introduction}
The ultracold OH molecule is a versatile platform for precision measurements. Its experimental appeal
lies in the fact that its ground $X^{2}\mathrm{\Pi}_{3/2}$ state is polar as well as paramagnetic,
and is therefore readily manipulated in the laboratory with the use of electric and magnetic fields \cite{Bochinski2003,Bochinski2004,Meerakker2005,Sawyer2007,Lemeshko2013}. Applications explored thus
far include precision spectroscopy \cite{Hudson2006,Kozlov2009} and quantum information processing
\cite{Lev2006}, in addition to studies of cold chemistry
\cite{Avdeenkov2002,Avdeenkov2003,Ticknor2005,Sawyer2008,Tscherbul2010} and quantum degeneracy
\cite{Quemener2012}.

In this article, we consider spin squeezing of the OH molecule. Our discussion occurs in the context
of the Heisenberg uncertainty relation \cite{Ma2011}
\begin{equation}
\label{eq:HUPJ}
\Delta J_{x}\Delta J_{y} \geq \left|\langle J_{z}\rangle \right|/2,
\end{equation}
between the three components of the angular momentum operator $J$. Spin squeezing refers to a situation
where the fluctuation in one of the components, say $\Delta J_{x}$, is reduced to below the standard
quantum limit $\sqrt{|\langle J_{z}\rangle|/2}$. Of course, the fluctuations in $J_{y}$ increase
correspondingly, in order to maintain the relation of Eq.~(\ref{eq:HUPJ}).

Spin squeezing constitutes a technique of interest at the frontiers of precision measurement and has
applications in spectroscopy, magnetometry, metrology, the detection of particle correlation and
entanglement, and quantum computation and simulation (\cite{Ma2011,Gross2012} and references therein).
The pioneering start to spin squeezing was provided by the work of Kitagawa and Ueda
\cite{Kitagawa1991,Kitagawa1993} and Wineland et al. \cite{Wineland1992,Wineland1994}, who considered
the squeezing of collective atomic spins, and was followed by many investigations
(\cite{GSA1989,Agarwal1990,Kuzmich1997,Hald1999,Sorensen1999,Kuzmich2000,Pu2000,Duan2000,Vernac2000,Law2001,
Raghavan2001,You2001,Vardi2001,Sorensen2001,Vernac2002,Andre2002,Berry2002,Sinha2003,Geremia2005,Choi2005,
Takeuchi2005,Yi2006,Jo2007,Fernholz2008,Esteve2008,You2008,Pezze2009,Takano2009,Gross2010,Leroux2010,Sau2010,
Benatti2011,Chen2011,Liu2011,Hamley2012,Diaz2012,Nemoto2014,Das2015,Opatrny2015}, for example). We emphasize that all
of this work relates to ensembles of correlated (pseudo)spins, represented by atoms (in a thermal
vapor or a degenerate gas), nuclei \cite{Rudner2011}, or molecules \cite{Hazzard2013}, where the
collective spin can assume a high value \cite{WineCom}.

More recently, squeezing has also been considered for single (atomic or nuclear) spins, or equivalently,
for an uncorrelated ensemble of such spins. Experiments have been carried out, using spin $3$
\cite{Soumya2007} and spin $7/2$ \cite{Auccaise2015} states of Cesium atoms. Possibilities also exist for
Dysprosium which offers states with spins from $8$ upto $12.5$. Theoretical calculations of single-spin
squeezing have been presented as well \cite{Rochester2012}. The squeezing in this case is limited by the
much smaller angular momentum available. Nonetheless, some effects can be experimentally relevant, for
example for precision spectroscopy \cite{Wineland1994}. Motivated by such a consideration, we
consider spin squeezing of a single OH molecule. Interesting applications to the measurement of magnetic
fields also seem possible \cite{Auzinsh2004,Cappellaro2009,Shah2010,Wasilewski2010,Sewell2012}, especially
following recent discussions of magnetometry using single SiC spins with angular momentum $3/2$ \cite{Lee2015};
ground state ultracold OH also carries rotational angular momentum $3/2$, and is sensitive to magnetic
fields via the Zeeman shift. As well, spin squeezing is related to the alignment-to-orientation transition
\cite{Rochester2012}, and we expect this perspective will be of relevance to our work, given the recent
interest in the stereochemical properties of the OH molecule \cite{Schmelcher2013}. We emphasize that
compared to previous proposals for spin squeezing molecules \cite{Hazzard2013}, which employ spin $1/2$
interacting molecules squeezed collectively, we consider a single noninteracting molecule with
angular momentum $3/2$.

In present-day laboratories, ultracold OH molecules are typically confined in magnetoelectrostatic traps
\cite{Meerakker2005,Sawyer2007,Lara2008,Stuhl2012,Quemener2012}. We therefore consider spin squeezing
enabled by these readily available static electric and magnetic fields. The advantage of using static,
rather than optical or microwave \cite{Ma2011}, fields for squeezing is that damping and decoherence due to
spontaneous emission can be avoided. Our starting point will be an effective eight-dimensional matrix
Hamiltonian that has been shown to model recent OH experiments quite well
\cite{Lara2008,Stuhl2012,Quemener2012}, and has also been diagonalized analytically \cite{Mishkat2013,Bohn2013}.
We demonstrate that for Zeeman and Stark shifts small compared to the Lambda-doublet splitting of the OH
ground state, this Hamiltonian can yield spin-squeezing of the types considered by Kitagawa and Ueda
\cite{Kitagawa1993}, Law et al. \cite{Law2001} and Agarwal and Puri \cite{GSA1989}. We provide analytic
and numerical results, discuss the optimization of the field strengths, and include the effects of field
misalignment in our treatment.

The remainder of this article is arranged as follows. Section~\ref{sec:HamDer} presents the derivation
of the spin squeezing Hamiltonian, Section~\ref{sec:Squeezing} discusses the dynamics of the squeezing parameter,
Section~\ref{sec:det} addresses the detection of the proposed squeezing, and Section~\ref{sec:Conc} supplies a
conclusion.

\section{Derivation of the spin squeezing Hamiltonian}
\label{sec:HamDer}
\subsection{The eight dimensional effective Hamiltonian}
Several experiments on ultracold OH molecules in crossed electric and magnetic fields, see Fig.1,
\begin{figure}[t]
\centering
\includegraphics[width=0.4\textwidth]{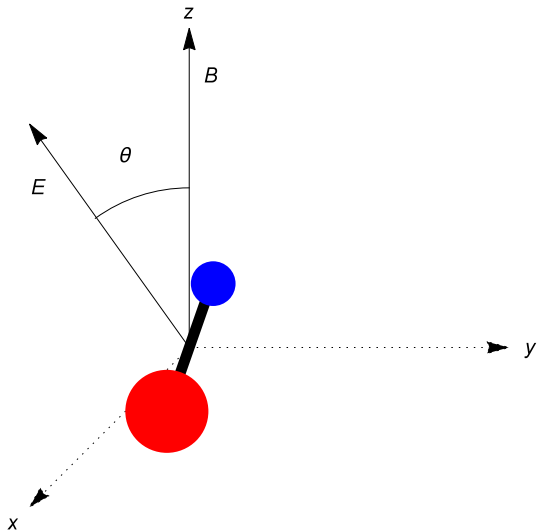}
\caption{(Color online) Schematic of the diatomic OH molecule in electric $E$ and magnetic $B$ fields
crossed at an angle $\theta$.}
\label{fig:P1}
\end{figure}
have been successfully modeled using an effective Hamiltonian involving only eight quantum states
\cite{Lara2008,Stuhl2012}. The domain of validity of this Hamiltonian and the details of the states involved
can be found in several articles \cite{Sawyer2007,Stuhl2012,Quemener2012,Bohn2013}, and will not be repeated
here. This effective Hamiltonian was recently diagonalized analytically following the detection of an underlying
chiral symmetry \cite{Mishkat2013}. In the process of identifying that symmetry, it was found that the OH
matrix Hamiltonian could be re-expressed in terms of two interacting spins (to avoid notational clutter we
set $\hbar = 1$; to make contact with laboratory parameters, requisite factors of $\hbar$ can be supplied to
any formula in this article by inspection, see the example supplied below),
\begin{equation}
\label{eq:OHHam}
H_{M}=-\tilde{\Delta}\sigma_{z}
-\tilde{B}J_{z}+\tilde{E}\sigma_{x}\left(J_{z}\cos\theta-J_{x}\sin\theta\right),
\end{equation}
where the constants are given by
\begin{equation}
\label{eq:cons}
\tilde{\Delta}=\frac{\Delta}{2}, \tilde{B}=\frac{4\mu_{B}B}{5}, \tilde{E}=\frac{2\mu_{e}E}{5},
\end{equation}
$\Delta$ being the Lambda-doublet splitting, $B$ and $E$ the uniform magnetic and electric fields, respectively,
$\mu_{B}$ the Bohr magneton and $\mu_{e}$ the electric dipole moment of the OH molecule. In
Eq.~(\ref{eq:OHHam}), $\theta$ is the angle between the electric and magnetic fields (see Fig.~\ref{fig:P1}), and
$\sigma=1/2$ and $J=3/2$ are the two interacting spins. The spin $\sigma$ is a pseudo-spin, with the spin
projections $\sigma_{z}=-1$ and $\sigma_{z}=1$ corresponding to the two Lambda-doublet manifolds of parity $e$ and $f$,
respectively. On the other hand, $J$ corresponds to the rotation of the molecular axis, and the values $J_{z}=\pm 1/2,\pm 3/2$
indicate the various projections of $J$ on the laboratory $z$ axis, which is chosen for convenience and without loss
of generality to be along the magnetic field. The matrix representation of $H_{M}$ is reproduced in the appendix for
the reader's convenience, see Eq.~(\ref{eq:Hmatrix}).

At this point the insertion of factors of $\hbar$ in Eq.~(\ref{eq:OHHam}) may be considered as a useful exercise.
We begin with the observation that $H_{M}$ has units of energy. Since the Pauli matrix $\sigma_{x}$ is dimensionless,
$\tilde{\Delta}$ needs to have units of energy, and thus must be multiplied by $\hbar$. On the other hand, $J_{z}$
has units of $\hbar$. Therefore $\tilde{B}$ needs to have units of frequency, and thus the right hand side of the
corresponding expression in Eq.~(\ref{eq:cons}) must be divided by an $\hbar$. Other formulas in this article can
be handled in a similar manner.

\subsection{The four dimensional squeezing Hamiltonian}
The generation of spin squeezing requires the presence of a nonlinearity, represented to lowest order by a term
quadratic in one of the spin operators in the relevant Hamiltonian \cite{Kitagawa1993}. However,
Eq.~(\ref{eq:OHHam}) is linear in each of the spin operators. It may not be readily obvious that the coupled
dynamics of the two spins can lead to squeezing, and it would assist our intuition if $H_{M}$ could be reduced,
even if in some restricted regime, to the form of a spin squeezing Hamiltonian familiar from the literature. In
order to effect such a reduction, we proceed by identifying a regime in which the pseudospin $1/2$ can be eliminated
adiabatically. As we show below, this results in an effective spin-squeezing Hamiltonian for the spin $3/2$ degree
of freedom. A similar procedure has been used earlier in the case of nuclear spin squeezing; in that case however
the eliminated spin is real \cite{Rudner2011}. The derivation begins with the Heisenberg equations implied by
Eq.~(\ref{eq:OHHam}) for the system variables
\begin{eqnarray}
\dot{\sigma}_{x}&=&2\tilde{\Delta}\sigma_{y},\label{eq:sx}\\
\dot{\sigma}_{y}&=&-2\tilde{\Delta}\sigma_{x}
-2\tilde{E}\sigma_{z}\left(J_{z}\cos\theta-J_{x}\sin\theta\right),\label{eq:sy}\\
\dot{\sigma}_{z}&=&2\tilde{E}\sigma_{y}\left(J_{z}\cos\theta-J_{x}\sin\theta\right),\label{eq:sz}\\
\dot{J_{x}}&=&\tilde{B}J_{y}-\tilde{E}\cos\theta \sigma_{x}J_{y},\label{eq:Jx}\\
\dot{J_{y}}&=&-\tilde{B}J_{x}+\tilde{E}\sigma_{x}\left(J_{x}\cos\theta+J_{z}\sin\theta\right),\label{eq:Jy}\\
\dot{J_{z}}&=&-\tilde{E}\sin\theta\sigma_{x}J_{y}.\label{eq:Jz}\\
\nonumber
\end{eqnarray}
We now consider the regime
\begin{equation}
\label{eq:adiacond}
\tilde{\Delta} > \tilde{E},\tilde{B},
\end{equation}
i.e. such that the Lambda-doublet splitting is larger than the Stark as well as Zeeman shifts. In our calculations
below we will work with $\tilde{E}/\tilde{\Delta} \simeq 0.25,$ which corresponds to an electric field of $100$ V/cm,
the lower bound stipulated by the value of stray fields currently affecting experiments \cite{Stuhl2012}. We will also
consider $\tilde{B}/\tilde{\Delta} \leq 0.1$, corresponding to a magnetic field of $20$ G, which can easily be
achieved experimentally.

In the regime indicated by Eq.~(\ref{eq:adiacond}), $\sigma_{x}$ and $\sigma_{y}$ will vary at the high rate
$\sim 2\tilde{\Delta}$ and may be thought of as the `fast' variables of our problem. We will see below that with
this assumption $\sigma_{z}$ ceases to be a dynamical variable, i.e. takes on a constant value. On the other hand,
$J_{x},J_{y}$ and $J_{z}$, varying at the low rates $\sim\tilde{B}$ and $\sim\tilde{E}$ (to first order),
may be thought of as the `slow' variables of the problem. We now adiabatically eliminate the fast spin variables
$\sigma_{x}$ and $\sigma_{y}$ in Eqs.~(\ref{eq:sx})-(\ref{eq:sy}) by setting the time derivatives in those equations
equal to zero. The resulting solutions are
\begin{eqnarray}
\sigma_{x}&=&-\frac{\tilde{E}C}{\tilde{\Delta}}\left(J_{z}\cos\theta-J_{x}\sin\theta\right),\label{eq:sxa}\\
\sigma_{y}&=&0,\label{eq:sya}\\
\sigma_{z}&=&C,\label{eq:sza}\\
\nonumber
\end{eqnarray}
where we have retained $C$ for generality. Note that the adiabatic solution $\sigma_{y}=0$ forces $\sigma_{z}$ to be a
constant [see Eq.~(\ref{eq:sz})], as mentioned above. We now use the solutions of Eqs.~(\ref{eq:sxa})-(\ref{eq:sza})
to adiabatically eliminate the fast spin $1/2$ degrees of freedom from Eq.~(\ref{eq:OHHam}). Dropping a constant term
proportional to $\tilde{\Delta},$ we obtain the adiabatic Hamiltonian for the spin $3/2$ variables
\begin{equation}
\label{eq:Hamad}
H_{a}=-\tilde{B}J_{z}-\frac{C\tilde{E}^{2}}{\tilde{\Delta}}\left(J_{z}\cos\theta-J_{x}\sin\theta\right)^{2},
\end{equation}
which clearly contains terms nonlinear in the angular momentum components, and can therefore enforce squeezing.

Below, we will show that some standard squeezing Hamiltonians can be recovered from Eq.~(\ref{eq:Hamad}), and
also explore the squeezing effects of this Hamiltonian for arbitrary $\tilde{B},\tilde{E}$ and $\theta$, within
the limits prescribed by the validity of the effective Hamiltonian [Eq.~(\ref{eq:OHHam})]. Subsequently, we will
compare these results with numerical calculations based on the full eight-dimensional Hamiltonian of
Eq.~(\ref{eq:OHHam}), which accounts for non-adiabatic effects and thus allows us to examine the validity of
the adiabatic elimination process used in arriving at Eq.~(\ref{eq:Hamad}). We note that Eq.~(\ref{eq:adiacond})
does not stipulate any relationship between $\tilde{E}$ and $\tilde{B}$, other than that they both have to be
smaller than $\tilde{\Delta}$. We will use this flexibility below to adjust the field magnitudes to optimize
squeezing.

\section{Spin squeezing dynamics}
\label{sec:Squeezing}
\subsection{Spin squeezing using the four-dimensional adiabatic Hamiltonian $H_{a}$}
In this section we consider spin squeezing using the four-dimensional adiabatic Hamiltonian
of Eq.~(\ref{eq:Hamad}). We provide analytic and numerical results, as appropriate.
\subsubsection{One-axis twisting}
\label{subsubsec:KU}
For $\tilde{B}=0,\theta=0, C=1$, Eq.~(\ref{eq:Hamad}) gives the Kitagawa-Ueda Hamiltonian \cite{Kitagawa1993},
\begin{equation}
H_{KU}=\tilde{\kappa}J_{z}^{2},
\end{equation}
where
\begin{equation}
\label{eq:tildekappa}
\tilde{\kappa}=-\frac{\tilde{E}^{2}}{\tilde{\Delta}}.
\end{equation}
The theoretical analysis for spin squeezing using $H_{KU}$ was first provided by Kitagawa and Ueda
\cite{Kitagawa1993}, and more recently by Rochester et al.\cite{Rochester2012}.
Nonetheless, we restate the procedure here, in order to compare with fully numerical calculations
to be presented later. The recipe for the analysis goes as follows. Using the four-dimensional matrix for
the $J=3/2$ operator $J_{z}$ \cite{SakuraiBook}, we write the matrix form for $H_{KU}$. We then
obtain the time evolution operator
\begin{equation}
U_{KU}=e^{-iH_{KU}t},
\end{equation}
which is also a four-dimensional matrix. We can then obtain the time evolution of any observable
$\mathcal{O}$ by using the relation $\mathcal{O}(t)=U_{KU}^{-1}\mathcal{O}U_{KU}.$ In this way we
find the matrix forms of the observables $J_{x}(t)$ and $J_{y}(t)$. In order to find the most
suitable axis for spin squeezing, it is convenient to consider a further rotation by an angle
$n$ about the $x$ axis \cite{Kitagawa1993}, i.e.
\begin{equation}
J_{y,n}(t)=e^{i n J_{x}(t)}J_{y}(t)e^{-i n J_{x}(t)},
\end{equation}
and so on for other observables. The operators $J_{x}(t)$ and $J_{x}^{2}(t)$ are unaffected by
this rotation, of course.

\begin{figure}[h]
\centering
\includegraphics[width=0.45\textwidth]{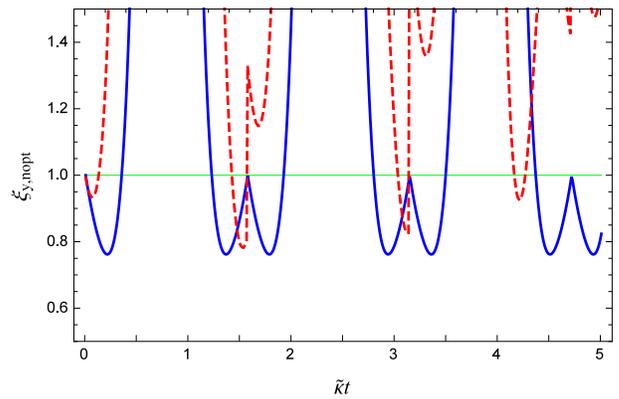}
\caption{(Color online) Plot of $\xi_{y,n_{\mathrm{opt}}}$ as a function of the dimensionless time $\tilde{\kappa}t$
where $\tilde{\kappa} $ is defined in Eq.~(\ref{eq:tildekappa}). The parameters are $\Delta=1.66$ GHz, $E=100$ V/cm
and $B=0,$ implying $\tilde{\kappa}=48$ MHz. Squeezing occurs when $\xi_{y,n_{\mathrm{opt}}}<1$. The blue (solid)
curve is the analytical prediction from Eqs.~(\ref{eq:nopt}) and (\ref{eq:WineSqueeze1}) based on the adiabatic
approximation of Eq.~(\ref{eq:adiacond}), with $\tilde{E}/\tilde{\Delta} \simeq 0.25$. The red (dashed) curve is
the numerical calculation using the eight dimensional effective Hamiltonian of Eq.~(\ref{eq:OHHam}). As can be seen,
nonadiabatic effects significantly change the magnitude and periodicity of the squeezing.}
\label{fig:P2}
\end{figure}

\begin{figure}[H]
\centering
\includegraphics[width=0.45\textwidth]{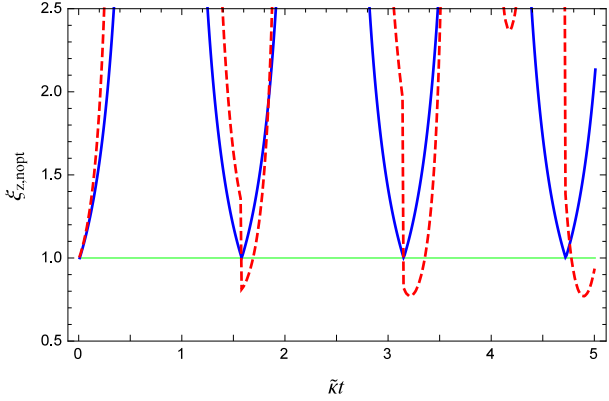}
\caption{(Color online) Plot of $\xi_{z,n_{\mathrm{opt}}}$ as a function of the dimensionless time $\tilde{\kappa}t$
where $\tilde{\kappa} $ is defined in Eq.~(\ref{eq:tildekappa}). The parameters are $\Delta=1.66$ GHz, $E=100$ V/cm
and $B=0,$ implying $\tilde{\kappa}=48$ MHz. Squeezing occurs when $\xi_{z,n_{\mathrm{opt}}}<1$. The blue (solid)
curve is the analytical prediction from Eqs.~(\ref{eq:nopt}) and (\ref{eq:WineSqueeze2}) based on the adiabatic
approximation of Eq.~(\ref{eq:adiacond}), with $\tilde{E}/\tilde{\Delta} \simeq 0.25$. The red (dashed) curve is
the numerical calculation using the eight dimensional effective Hamiltonian of Eq.~(\ref{eq:OHHam}). As can be seen,
nonadiabatic effects actually introduce some squeezing into the otherwise `anti-squeezed' quadrature.}
\label{fig:P3}
\end{figure}

For our initial state, we choose, following Kitagawa and Ueda, the (coherent) stretched state
along the $x$ axis \cite{Kitagawa1993},
\begin{equation}
\ket{i}_{KU}=\ket{J = \frac{3}{2}, M = \frac{3}{2}}_{\hat{x}}=2^{-3/2}\sum_{k=0}^{3} {{3}\choose{k}}
\ket{\frac{3}{2}, \frac{3}{2}-k},
\end{equation}
which we have expanded in the $z$ basis on the right hand side. We note that the current level of control
over the OH ground state manifold should readily allow this state to be prepared \cite{Stuhl2012}. Using
the initial state $\ket{i}_{KU}$ we find the expectation values
$\langle J_{y,n}(t)\rangle, \langle J_{y,n}^{2}(t)\rangle$, etc., and thence the variances such as
\begin{equation}
\left(\Delta J_{y,n}(t)\right)^{2}=\langle J_{y,n}^{2}(t)\rangle-\langle J_{y,n}(t)\rangle^{2}.
\end{equation}
This procedure yields the relevant quantities \cite{Rochester2012}
\begin{eqnarray}
\langle J_{x}(t)\rangle & = & \frac{3}{2}\cos^{2}\tilde{\kappa}t,\\
\left(\Delta J_{y,n}(t)\right)^{2} & = &\frac{3}{4}
\left[1+\frac{M}{2}+\frac{\sqrt{M^{2}+N^{2}}}{2}\cos \left(2n+2\delta\right)\right],\nonumber\\
\left(\Delta J_{z,n}(t)\right)^{2} & = &\frac{3}{4}
\left[1+\frac{M}{2}-\frac{\sqrt{M^{2}+N^{2}}}{2}\cos \left(2n+2\delta\right)\right],\nonumber\\
\end{eqnarray}
where
\begin{equation}
M=1-\cos 2\tilde{\kappa}t,\,\,N=2\sin 2\tilde{\kappa}t, \,\,\delta=\frac{1}{2}\tan^{-1}\left(\frac{N}{M}\right).
\end{equation}
As pointed out earlier \cite{Kitagawa1993,Rochester2012}, the $y$ quadrature is maximally squeezed for $\cos \left(2n_{\mathrm{opt}}+2\delta\right)=-1$, i.e. along the axis
\begin{equation}
\label{eq:nopt}
n_{\mathrm{opt}}=\frac{\pi}{2}-\delta,
\end{equation}
in the $y-z$ plane. The amount of squeezing is quantified by the parameter introduced by Wineland et al.
\cite{Wineland1994}
\begin{equation}
\label{eq:WineSqueeze1}
\xi_{y,n}=\sqrt{3}\frac{\langle \left(\Delta J_{y,n}(t)\right)\rangle}{\left|\langle J_{x}(t)\rangle \right|},
\end{equation}
and, analogously,
\begin{equation}
\label{eq:WineSqueeze2}
\xi_{z,n}=\sqrt{3}\frac{\langle \left(\Delta J_{z,n}(t)\right)\rangle}{\left|\langle J_{x}(t)\rangle \right|}.
\end{equation}
Squeezing occurs when either $\xi_{y,n}$ or $ \xi_{z,n}$ is less than one. We have shown a plot of
$\xi_{y,n_{\mathrm{opt}}}$ in Fig.~\ref{fig:P2}. The blue (solid) line is the analytical prediction of
Eq.~(\ref{eq:WineSqueeze1}) and shows that the $y$ quadrature is squeezed periodically. As found by
Rochester et al. \cite{Rochester2012}, for $J=3/2$ the minimum value of the squeezing parameter is
m$\xi^{\mathrm{min}}_{y,n_{\mathrm{opt}}} \simeq 0.75$, consistent with Fig.~\ref{fig:P2}.
The red (dashed) curve in Fig.~\ref{fig:P2} corresponds to a numerical calculation of squeezing using the
full effective Hamiltonian of Eq.~(\ref{eq:OHHam}), which incorporates the effects of nonadiabaticity, see
Section \ref{subsec:eightd} below. As can be seen, nonadiabatic effects change the magnitude and periodicity
of squeezing quite significantly. This may be expected, as our adiabaticity parameter
$\tilde{E}/\tilde{\Delta}\sim 0.25$ is quite large. A plot of  $\xi_{z,n_{\mathrm{opt}}}$ has been shown in
Fig.~\ref{fig:P3}. The analytical prediction of Eq.~(\ref{eq:WineSqueeze2}), represented by the blue (solid)
curve, implies that the shown $z$ quadrature is never squeezed. Interestingly, the inclusion of nonadiabatic
effects, shown by the red (dashed) curve, introduces some squeezing of this quadrature.

\subsubsection{Uniform field Hamiltonian}
\label{subsubsec:LNL}
In this section we consider spin squeezing in the presence of a magnetic field. For $\tilde{B}\neq 0$,
and $\theta=\pi/2$, Eq.~(\ref{eq:Hamad}) takes the form
\begin{equation}
\label{eq:HLNL}
H_{LNL}=-\tilde{B}J_{z}+\tilde{\kappa}J_{x}^{2}.
\end{equation}
This Hamiltonian is, to within a unitary anti-clockwise rotation of $\pi/2$ around the $y$ axis, the same as
proposed earlier by Law, Ng and Leung \cite{Law2001,Rojo2003,Vidal2004,Jin2007}. These authors suggested adding the uniform
field term to the Kitagawa-Ueda Hamiltonian as it yielded greater squeezing for longer times. Analytic solutions
to the spin squeezing dynamics of Eq.~(\ref{eq:HLNL}) are not available for arbitrary spin $J$. Generally, analytic
results can be found only for $J \leq 2$, since the eigenvalues of the Hamiltonian are required for the calculation,
which can be found in closed form only for matrices of dimension $5$ or lower. Some results for the case where $J=3/2$
is a collective spin have been published in the literature \cite{Kim2007,Jin2007}. We provide additional expressions
in order to discuss the details of our problem.

\begin{figure*}[t]
\centering
\includegraphics[width=0.95\textwidth]{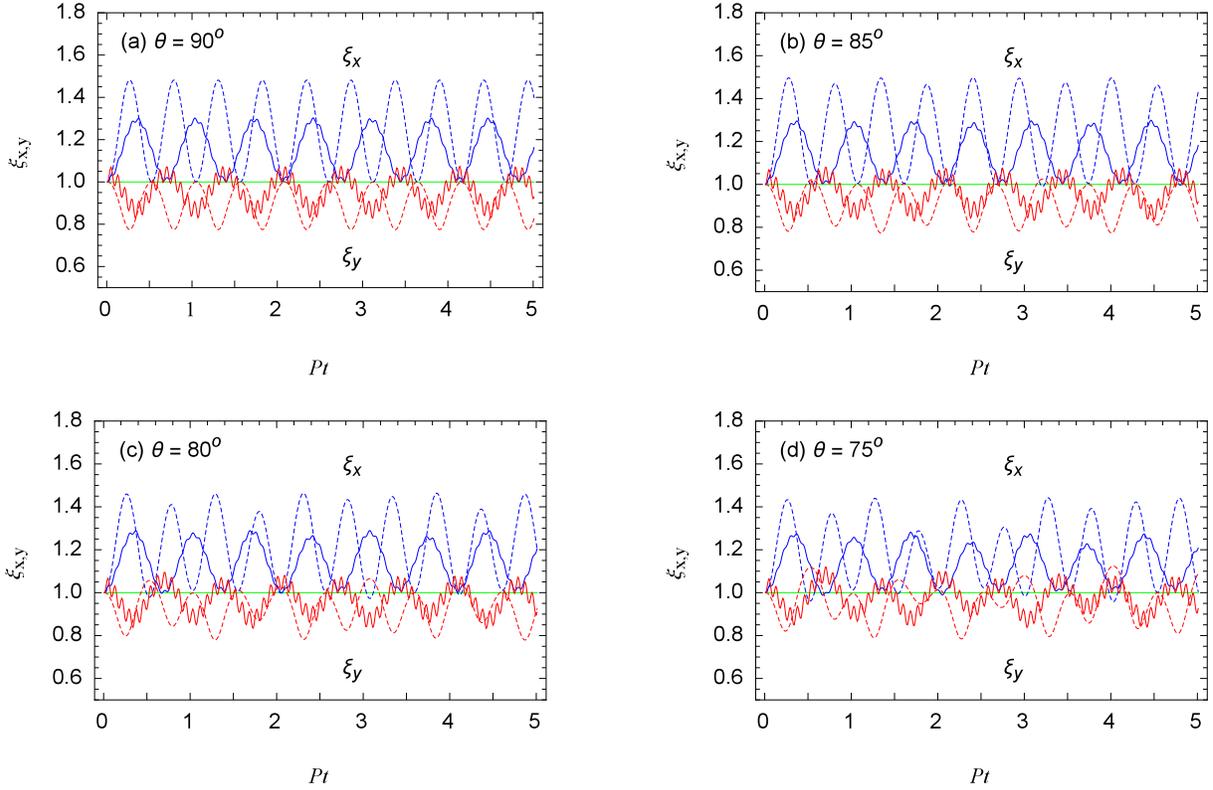}
\caption{(Color online) Plots of $\xi_{x,y}$ as functions of the dimensionless time $Pt$
where $P$ is defined in Eq.~(\ref{eq:Pdef}). The parameters are $\Delta=1.66$ GHz, $E=100$ V/cm, $B=20$ G,
implying $P=144$ MHz and (a) $\theta =90^\circ$ (b) $\theta=85^\circ$ (c) $\theta=80^\circ$ (d) $\theta=75^\circ$.
Squeezing occurs when $\xi_{x,y}<1$. The dashed curves are the analytical predictions for $\xi_{x}$
[Eq.~(\ref{eq:xixLNL})] (blue) and $\xi_{y}$ [Eq.~(\ref{eq:xiyLNL})] (red). The solid curves represent
numerical calculations carried out using the eight dimensional effective Hamiltonian of Eq.~(\ref{eq:OHHam}).
As can be seen by comparing with Fig.~\ref{fig:P2}, the squeezing is quite robust to nonadiabatic effects. Also,
deviations of about $5^\circ$ away from the nominal value of $90^\circ$ for field alignment do not affect the
squeezing greatly.}
\label{fig:P4}
\end{figure*}

\begin{figure}[h]
\centering
\includegraphics[width=0.45\textwidth]{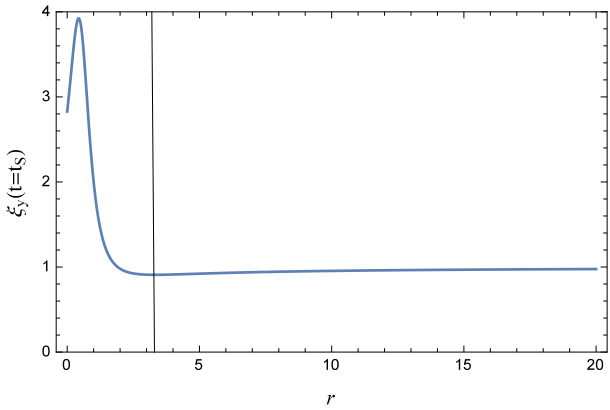}
\caption{(Color online) Plot of $\xi_{y}(t=t_{S})$ [Eq.~(\ref{eq:xiyts})] versus $r$ [Eq.~(\ref{eq:rval})]. The
vertical line corresponds to $r \simeq 3.3$, at which $\xi_{y}(t=t_{S})$ has a minimum, corresponding to maximum
squeezing.}
\label{fig:P5}
\end{figure}

To determine the squeezing, we follow a procedure similar to that of Section~\ref{subsubsec:KU}, but consider
the initial stretched state along the $z$ direction
\begin{equation}
\ket{i}_{LNL}=\ket{0,0,0,1},
\end{equation}
and determine the squeezing about the $x$ and $y$ axes. We find
\begin{eqnarray}
\langle J_{x}(t)\rangle &= & \langle J_{y}(t)\rangle =0,
\label{eq:JxytLNL}\\
\langle J_{z}(t)\rangle & = & \frac{3}{2}\left[1-\left(\frac{\tilde{\kappa}}{P}\sin P t\right)^{2}\right],\\
\langle \left(\Delta J_{x}(t)\right)\rangle^{2} & = &\frac{3}{4}
\left[1+2\tilde{\kappa}\tilde{B}\left(\frac{\sin P t}{P}\right)^{2}\right],\\
\langle \left(\Delta J_{y}(t)\right)\rangle^{2} & = &\frac{3}{4}
\left[1-2\tilde{\kappa}\left(\tilde{B}-\tilde{\kappa}\right)\left(\frac{\sin P t}{P}\right)^{2}\right],
\label{eq:DJytLNL}\\
\nonumber
\end{eqnarray}
where
\begin{equation}
\label{eq:Pdef}
P=\sqrt{\tilde{B}^{2}-\tilde{B}\tilde{\kappa}+\tilde{\kappa}^{2}}.
\end{equation}
Using Eqs.~(\ref{eq:JxytLNL})-(\ref{eq:DJytLNL}), we can find the squeezing parameters
\begin{equation}
\label{eq:xixLNL}
\xi_{x}=\sqrt{3}\frac{\langle \left(\Delta J_{x}(t)\right)\rangle}{\left|\langle J_{z}(t)\rangle \right|},
\end{equation}
and
\begin{equation}
\label{eq:xiyLNL}
\xi_{y}=\sqrt{3}\frac{\langle \left(\Delta J_{y}(t)\right)\rangle}{\left|\langle J_{z}(t)\rangle \right|},
\end{equation}
Plots of the squeezing dynamics are shown in Fig.~\ref{fig:P4}(a) for the same parameters as in
Fig.~(\ref{fig:P2}) but with $B=20$ G. The dashed curves represent the analytical results for $\xi_{x}$
[Eqs.~(\ref{eq:xixLNL})] (blue) and $\xi_{y}$ [Eq.~(\ref{eq:xiyLNL})] (red). The solid curves are the
corresponding numerical calculations using $H_{M}$ from Eq.~(\ref{eq:OHHam}).
It can be seen from the numerically calculated red (solid) curves in Fig.~\ref{fig:P4}(a)
that the Law-Ng-Leung approach is quite robust to nonadiabatic effects. In  Fig.~\ref{fig:P4} we have chosen
a magnetic field optimized using the following procedure. From the analytic result of Eq.~(\ref{eq:xiyLNL}), it
can be seen that at multiples of the time $t_{S}$
\begin{equation}
t_{S} = \frac{\pi}{4P},
\end{equation}
the squeezing parameter for the $y$ quadrature attains an extremum value given by
\begin{equation}
\label{eq:xiyts}
\xi_{y}(t=t_{S})=\frac{2\sqrt{\left(r^{2}-r+1\right)\left(r^{2}-2r+2\right)}}{2r^{2}-2r+1},
\end{equation}
where the dimensionless ratio
\begin{equation}
\label{eq:rval}
r = \frac{\tilde{B}}{\left|\tilde{\kappa}\right|}.
\end{equation}
The denominator of $\xi_{y}(t=t_{S})$ vanishes only for the complex values $r=(1\pm i)/2$, which are
excluded by experiment. Thus, $\xi_{y}^{\mathrm{min}}$ stays finite as $r$ varies, as can be seen from
Fig.~\ref{fig:P5}.
Differentiation of Eq.~(\ref{eq:xiyts}) readily yields a minimum value $\xi_{y}^{\mathrm{min}}(t=t_{S})\simeq 0.8$,
which occurs at $r \simeq 3.3$, indicated by the vertical line in Fig.~\ref{fig:P5}. This optimized value of $r$
corresponds to the magnetic field $20$ G used in Fig.~\ref{fig:P4}(a). For simplicity, unlike in Ref.~\cite{Jin2007},
we have not found the time-dependent axis of optimum squeezing, which can lead to even better squeezing than
we have presented.

\subsubsection{General case}
\label{subsubsec:AnyTheta}
In this section we consider spin squeezing as a function of the angle between the electric and magnetic fields.
The interest in this degree of freedom arises from the necessity of accounting for possible misalignments between
the electric and magnetic fields in the laboratory, e.g. away from the nominal value of $\theta=\pi/2$ for $H_{LNL}$,
see Section~\ref{subsubsec:LNL}. While this case can be solved analytically as well, the expressions are lengthy and
we provide numerical solutions instead. The Hamiltonian is
\begin{equation}
\label{eq:Hg}
H_{g}=-\tilde{B}J_{z}+\tilde{\kappa}\left(J_{z}\cos\theta-J_{x}\sin\theta\right)^{2}.
\end{equation}
Before we proceed further, we mention that this Hamiltonian, when rotated anti-clockwise
by an angle $\theta$ about the $y$ axis yields
\begin{equation}
H_{g}'=e^{i\theta J_{y}}H_{g}e^{-i\theta J_{y}}=-\tilde{B}\left(J_{z}\cos\theta-J_{x}\sin\theta\right)+\tilde{\kappa}J_{z}^{2},
\end{equation}
which is of the form considered earlier by Agarwal and Puri for arbitrary $J$ \cite{GSA1989}.

Numerical plots are presented in Fig.~\ref{fig:P4} for (b) $\theta=85^\circ$ (c) $\theta=80^\circ$ and (d)
$\theta=75^\circ$, respectively. The dashed curves are the numerical implications of Eq.~(\ref{eq:Hg}) for
$\xi_{x}$ [Eq.~(\ref{eq:xixLNL})] (blue) and $\xi_{y}$ [Eq.~(\ref{eq:xiyLNL})] (red). The solid curves are the
corresponding numerical calculations starting from Eq.~(\ref{eq:OHHam}). As can be seen, the squeezing is quite
robust to field misalignment. Only at about $\theta = 75^\circ$ does the pattern change noticeably from that at
$\theta=90^\circ$. Interestingly, while misalignment degrades squeezing in the $y$ quadrature, it does not
correspondingly introduce squeezing in the $x$ quadrature.

\subsection{Spin squeezing using the eight-dimensional effective Hamiltonian $H_{M}$}
\label{subsec:eightd}
In this section we describe our spin squeezing calculations using the full eight-dimensional Hamiltonian
$H_{M}$ of Eq.~(\ref{eq:OHHam}). This enables us to quantitatively examine the adiabatic approximation made in
deriving the four-dimensional spin squeezing Hamiltonian $H_{a}$ of Eq.~(\ref{eq:Hamad}). The calculation
proceeds as follows. We begin with the eight-dimensional matrix representation of $H_{M}$ from Eq.~(\ref{eq:Hmatrix}).
Although the treatment can be carried out analytically, the expressions are very long, and we calculate instead
numerically the time evolution operator
\begin{equation}
U_{M}=e^{-i H_{M}t},
\end{equation}
which is also an eight-dimensional matrix. Starting from the initial state $\ket{\psi (0)}$, represented by
an eight-dimensional column matrix, the state vector $\ket{\psi (t)}$ at any later time $t$ is
\begin{equation}
\ket{\psi (t)}=U_{M}\ket{\psi (0)}.
\end{equation}
The density matrix of the full system can then be easily foun
\begin{equation}
\rho (t)= \ket{\psi (t)}\bra{\psi (t)}.
\end{equation}
The reduced density matrix of the spin $3/2$ system can then be found by tracing over the spin $1/2$ degrees
of freedom, i.e.
\begin{equation}
\rho_{3/2}(t)= \mathrm{Tr}_{1/2}\left[\rho(t)\right].
\end{equation}
From here the expectation values of the relevant operators can be calculated, such as
\begin{equation}
\langle J_{x}(t)\rangle= \mathrm{Tr}_{3/2}\left[J_{x}\rho_{3/2}(t)\right],
\end{equation}
and therefore also the squeezing parameters, as for example in Eq.~(\ref{eq:xixLNL}). It may be
useful to mention that for Section~\ref{subsubsec:KU} the initial state is now written as
\begin{equation}
\ket{i}_{KU}=\frac{1}{2\sqrt{2}}\ket{0,0,0,0,1,\sqrt{3},\sqrt{3},1},
\end{equation}
and for Sections~\ref{subsubsec:LNL} and \ref{subsubsec:AnyTheta} as
\begin{equation}
\ket{i}_{LNL}=\ket{0,0,0,0,0,0,0,1}.
\end{equation}
Comparison of the four and eight-dimensional results can be seen in Figs.~\ref{fig:P2}-\ref{fig:P4},
and has been described in the preceding sections. These plots imply that while non-adiabatic effects
can be important, the adiabatic approximation captures the basic spin dynamics reasonably well, and
the intuition afforded by this approximation about the existence of squeezing is well founded. Better
agreement can be produced by assuming a smaller adiabaticity parameter, but currently this is limited
by stray electric fields in the experiments. The key point is that even in the presence of
non-adiabaticity, squeezing exists.

\section{Detection of spin squeezing}
\label{sec:det}
The proposed spin squeezing may be detected by performing quantum tomography on the OH ground state
manifold, which would yield the density matrix, from which squeezing information can readily be
extracted. Such procedures can be carried out for OH in analogy to experiments performed earlier on
atomic \cite{Soumya2007} and nuclear \cite{Auccaise2015} systems. In the laboratory, the predicted
squeezing will be degraded by damping and noise, due to molecular collisions and trap loss. In the
present work, we have justifiably neglected these effects, since they occur at typical rates of Hz
(collisions) \cite{Avdeenkov2002,Meerakker2005,Sawyer2007} or KHz (trap loss) \cite{Stuhl2012}, while
squeezing is generated at frequencies of MHz [Fig.~\ref{fig:P2}].

\section{Conclusion}
\label{sec:Conc}
We have proposed a scheme for spin squeezing the ultracold OH molecule, in the context of ongoing
experiments. Production of such nonclassical states is expected to be useful for spectroscopy,
magnetometry and stereochemistry. We have identified an accessible parameter regime where single-axis
twisting as well as uniform field squeezing can be implemented with the use of static fields only. Since
we do not propose to use optical or microwave fields, our scheme is free from damping and decoherence
due to spontaneous emission and optical pumping. In our analysis, we have shown how to optimize the field
values and also investigated the effect of field misalignment on the squeezing. To our knowledge, our work
is the first concrete proposal for spin squeezing single noninteracting molecules with angular momentum
greater than $1/2$. We note that with the use of additional electric and magnetic fields, other spin
squeezing Hamiltonians may also be realized using OH, such as the one proposed by Raghavan et al.
\cite{Raghavan2001}. Also, a more accurate description of the OH ground state can be reached by including
more states in the Hamiltonian, accounting for fine and hyperfine structure and electric quadrupole
interactions \cite{Sawyer2007,Maeda2015}. It would be interesting to investigate the effect of these
additions on our results. Finally, our scheme can also be extended to the ground states of other polar
paramagnetic molecules such as $^{2}\Pi_{3/2}$ LiO and $^{3}\Phi_{2}$ CeO.

We would like to thank S. Marin for useful discussions.

\section{Appendix}
In this Appendix we provide, for the reader's convenience, the matrix representation of $H_{M}$ from
Eq.~(\ref{eq:OHHam}) \cite{Lara2008,Mishkat2013},
\begin{widetext}
\begin{equation}
\label{eq:Hmatrix} H_{M} = \\
\small
\begin{pmatrix}
  -\frac{\hbar\Delta}{2}-\frac{6}{5}\mu_{B}B      &     0         &    0        &       0 & \frac{3}{5}\mu_{e}E\cos\theta   & -\frac{\sqrt{3}}{5}\mu_{e}E\sin\theta            &    0         & 0  \\
           0     & -\frac{\hbar\Delta}{2}-\frac{2}{5}\mu_{B}B    & 0    &      0 & -\frac{\sqrt{3}}{5}\mu_{e}E\sin\theta  & \frac{1}{5}\mu_{e}E\cos\theta            &  -\frac{2}{5}\mu_{e}E\sin\theta           &  0 \\
           0     & 0       &  -\frac{\hbar\Delta}{2}+\frac{2}{5}\mu_{B}B   &   0  &  0      &   -\frac{2}{5}\mu_{e}E\sin\theta            &             -\frac{1}{5}\mu_{e}E\cos\theta       & -\frac{\sqrt{3}}{5}\mu_{e}E\sin\theta  \\
           0&0&0&-\frac{\hbar\Delta}{2}+\frac{6}{5}\mu_{B}B &0 &0 &-\frac{\sqrt{3}}{5}\mu_{e}E\sin\theta &-\frac{3}{5}\mu_{e}E\cos\theta \\
\frac{3}{5}\mu_{e}E\cos\theta  & -\frac{\sqrt{3}}{5}\mu_{e}E\sin\theta  &       0        & 0  & \frac{\hbar\Delta}{2}-\frac{6}{5}\mu_{B}B  &             0&     0        &  0 \\
-\frac{\sqrt{3}}{5}\mu_{e}E\sin\theta   & \frac{1}{5}\mu_{e}E\cos\theta & -\frac{2}{5}\mu_{e}E\sin\theta & 0 & 0  & \frac{\hbar\Delta}{2}-\frac{2}{5}\mu_{B}B             &   0          &  0 \\
       0    &  -\frac{2}{5}\mu_{e}E\sin\theta & -\frac{1}{5}\mu_{e}E\cos\theta  & -\frac{\sqrt{3}}{5}\mu_{e}E\sin\theta & 0 & 0   &\frac{\hbar\Delta}{2}+\frac{2}{5}\mu_{B}B   & 0 \\
       0    &  0     & -\frac{\sqrt{3}}{5}\mu_{e}E\sin\theta  & -\frac{3}{5}\mu_{e}E\cos\theta &   0  & 0 &0 &\frac{\hbar\Delta}{2}+\frac{6}{5}\mu_{B}B   \\
\end{pmatrix}.
\end{equation}
\end{widetext}


\begin{thebibliography}{27}
\bibitem[{wco()}]{WineCom}
\bibinfo{note}{We note however that the possibility of squeezing a single  atomic spin with $J > 1/2$
was suggested by Wineland et al. \cite{Wineland1992,Wineland1994}.}
\end{thebibliography}
\end{document}